\documentclass[12pt]{article}
\usepackage{graphicx} % Required for inserting images
\usepackage{authblk} % Cleaner for affiliations
\usepackage[margin=1in]{geometry} % Adjust page margins
\usepackage[round]{natbib} % add package for citations
\usepackage{xurl} % helpful for urls 
\usepackage{multicol} % for appendix
\usepackage{amsmath} % for equations

\title{Are Betting Markets Better than Polling in Predicting Political Elections?}

\author[1]{Laurie E. Cutting}
\author[2]{Sarah S. Hughes-Berheim}
\author[3]{Paul M. Johnson}
\author[4]{Hiba Baroud}
\author[5]{Brett Goldstein}

\affil[1]{Peabody College, Vanderbilt University, Institute of National Security}
\affil[2]{Peabody College, Vanderbilt University}
\affil[3]{Civil and Environmental Engineering, Vanderbilt University}
\affil[4]{Civil and Environmental Engineering, Vanderbilt University}
\affil[5]{School of Engineering, Vanderbilt University, Institute of National Security}

\date{July 07, 2025}

\begin{document}

\maketitle

\begin{abstract}
    Political elections are one of the most significant aspects of what constitutes the fabric of the United States. In recent history, typical polling estimates have largely lacked precision in predicting election outcomes, which has not only caused uncertainty for American voters, but has also impacted campaign strategies, spending, and fundraising efforts. One intriguing aspect of traditional polling is the types of questions that are asked – the questions largely focus on asking individuals who they intend to vote for. However, they don’t always probe who voters think will win – regardless of who they want to win. In contrast, online betting markets allow individuals to wager money on who they expect to win, which may capture who individuals think will win in an especially salient manner. The current study used both descriptive and predictive analytics to determine whether data from Polymarket, the world’s largest online betting market, provided insights that differed from traditional presidential polling. Overall, findings suggest that Polymarket was superior to polling in predicting the outcome of the 2024 presidential election, particularly in swing states. Results are in alignment with research on “Wisdom of Crowds” theory, which suggests a large group of people are often accurate in predicting outcomes, even if they are not necessarily experts or closely aligned with the issue at hand. Overall, our results suggest that betting markets, such as Polymarket, could be employed to predict presidential elections and/or other real-world events. However, future investigations are needed to fully unpack and understand the current study’s intriguing results, including alignment with Wisdom of Crowds theory and portability to other events.
\end{abstract}

\section{Introduction}
Political elections are one of the most significant aspects of what constitutes the fabric of the United States. As such, being able to quantitatively predict who might win is a vast industry. For example, about \$4.7 billion dollars were raised just for the 2024 presidential election cycle \citep{baileyNewParadigmPolling2023}. This large investment is understandable: being able to predict the future, especially for something as critical as a political election, has profound economic, world, and social implications both in the short and long-term \citep{guruEconomicImpactUnited2024, marmorPresidentialElectionUS2004}. Therefore, political campaigns devote a substantial portion of money raised towards polling – estimates suggest nearly 10\% of the total \$4.7 billion dollars raised (or \$50 million) was spent on polling in the 2024 election cycle \citep{jeffrey-wilensky16BillionWill2024,learnerDonaldTrumpKamala2024}.

\subsection{History of Polling}

While present day elections rely significantly on polling, the reliance on polling approaches to guide election strategy is not new (for a detailed review see \citep{geerPublicOpinionPolling2004}). Indeed, the history of polling begins several centuries ago, in 1824, with the first recorded straw poll suggesting President Andrew Jackson would defeat John Quincy Adams to be the 7th president of the United States \citep{tankardjrPublicOpinionPolling1972}. After its conception, election polling increased in popularity, and with it, more sophisticated methods of polling were developed. These methods included developing approaches for random sampling \citep{crossleyStrawPolls19361937} as well as capturing demographic characteristics, such as political party, sex, and religion \citep{friedGoverningPolls2010}. As such, the first nationwide presidential polls, conducted in the 1930s \citep{gallupGallupPollPublic1972,literarydigestFirstVotesDigests1936}, remain relatively similar to polling methods used in the present day: people are queried through some mode of communication and asked the quintessential questions: “Do you plan to vote? and If so, who do you plan to vote for?”. From these data, polling companies provide valuable information to political campaigns and to the public.

Although the goal of polling – to query future outcomes - hasn’t changed much across time, various modalities for querying individuals have emerged with the advent of new technologies (e.g., phones, texting, internet), which has resulted in more diverse forms of polling, as well as more sophisticated sampling methods. In 2002, the Help America Vote Act required states to generate databases containing information about a voter’s name, address, voting history, and political party \citep{hillygusEvolutionElectionPolling2011}. At a similar time, telephone polling was ramping up, and polls had switched from random digit dialing to using these databases to target registered voters by their home and/or cell phone \citep{hillygusPersuadableVoterWedge2014}. This type of polling continued to increase until 2012, after which it began to decline \citep{kennedyHowPublicPolling2023}, likely due to a continuing trend of people not answering their phones \citep{kennedyResponseRatesTelephone2019}. Instead, online polling and texting emerged as popular polling modalities due to their ability to reach more people \citep{kennedyHowPublicPolling2023,mercerComparingTwoTypes2023}, thus improving estimation accuracy. These methods also allowed for increases in both probability and non-probability based (Opt-In) polling, which were common methods during the 2016, 2020, and 2024 elections \citep{kennedyHowPublicPolling2023}.

\subsection{Recent Polling Inaccuracies}

Despite these advancements, polling approaches have not always been successful in predicting outcomes. In recent history, typical polling estimates have largely lacked precision in predicting election outcomes, especially for the 2016 and 2020 presidential elections \citep{clintonAAPORTaskForce2021, campbellMisfiresSurprisesPolling2022, kennedyEvaluation2016Election2018}. These outcomes left Americans confused and pollsters scrambling to make adjustments to improve polling accuracy. Indeed, after the 2016 election, in which the polls incorrectly predicted Hilary Clinton to be the 45th president of the United States, there was a large push to increase the amount of online polling being done \citep{mortimorePollsTheirContext2017} due to concerns about social desirability biases, cost effectiveness, and data efficiency \citep{hargittaiBiasesOnlinePolitical2018, jenningsElectionPollingErrors2018}. These pushes were especially prevalent within key swing states where the polling estimates were the most inaccurate \citep{kennedyHowPublicPolling2023}. During this online polling push, opt-in polling, in which a large percentage of online users were given the opportunity to “opt-in” or choose to participate, also increased in popularity \citep{kennedyHowPublicPolling2023}. This type of non-probabilistic, convenience sampling may have resulted in a non-representative sample that skewed the results of the 2020 election polls \citep{goldfarbGettingItRight2013}– which suggested a much closer race between the projected winner, President Biden, and Donald Trump than what was observed (President Biden won by a 4-point margin) \citep{igielnikBidens2020Victory2021}. For a detailed review and evaluation of the 2020 general election polls see \citet{clintonAAPORTaskForce2021}. 

In response to the failures of the 2020 polls, pollsters implemented several methodological changes to improve their polling strategies. These included the key changes of reaching participants via texting and ending “opt-in” polling \citep{kennedyHowPublicPolling2023}. As a result, there was a rise in probability-based panels, for which a group of people are randomly selected via their address to participate in polling surveys, which have greater accuracy over non-probability sampling (Opt-in polling) \citep{mercerComparingTwoTypes2023, mercerOnlineOptinPolls2024}. These changes led to improved accuracy in the 2024 polls compared to both 2016 and 2020 \citep{murrayWhat2024Polls2024}; however, in general, the polls still overestimated Kamala Harris’s chances of winning and underestimated Donald Trump’s, especially within swing states. These results only further contributed to the public’s continued skepticism of pre-election polls.

\subsection{The Importance of Polling Accuracy}

Inaccurate pre-election polls not only cause uncertainty for American voters, but they can also impact campaign strategies, spending, and fundraising efforts. For example, many presidential candidates use polling data to determine areas in the country where their messaging strategy can be improved (Jacobs and Shapiro 2005). Once these areas are identified, campaigns spend additional resources, typically via online and social media advertising \citep{moserHowUSPresidential2024}, to improve their standings in these areas and increase their overall winning likelihood. Inaccurate polling data likely results in campaigns incorrectly allocating millions of dollars (or more) to improve their standings in already secured areas or overestimating their messaging strategy amongst a key demographic, therefore not spending critical resources where necessary. Further, polling data directly impacts fundraising efforts, which is positively associated with electoral success \citep{leMoneyPoliticsHow2024}. Indeed, research shows positive polling numbers are associated with increased fundraising efforts \citep{mostafaviTaleTwoMetrics2021}, suggesting big donors and others use polls to decide which presidential candidate to financially support and when \citep{koerthHowMoneyAffects2018}.

These examples underscore the importance of collecting accurate presidential polling data both for the American people, who use polls to plan for the future and gauge American values \citep{olssonHarvestingWisdomCrowds2019}, and for campaigns, who use polls to determine where and when financial investments should be made \citep{jacobsPollingPoliticsMedia2005}. As such, the inaccuracy of presidential polling data within the last decade has led many to question the future of polling data in American elections. Several recent articles cite new techniques for distributing, collecting, and analyzing polling data \citep{baileyNewParadigmPolling2023, lauderdaleModelbasedPreelectionPolling2020, levyAnalyzingEffectRegional2023}, all aimed at improving polling accuracy. Yet, it remains uncertain whether these methods will be widely adopted, and if so, whether they will address the persistent issues in polling.

\subsection{A Different Approach: Online Betting Markets}

One intriguing aspect of traditional polling is the types of questions that are asked. The questions are largely focused on asking individuals their opinions about candidates on various issues, as well as who they intend to vote for. Interestingly, they don’t always probe voters on who they think will win – regardless of who they want to win \citep{hulettElectionsInsightsWhy2024, olssonHarvestingWisdomCrowds2019}. For example, the 2024 Atlas Election Poll, a select pollster awarded for its reliability \citep{rogersThreeThingsLook2024}, only reported results of who their participants (1) planned to vote for and (2) would vote for in a head-to-head race between Trump and Harris. While who a voter thinks will win versus who they want to win can be congruent, they aren’t necessarily. This is a subtle but critical distinction: one can intensely dislike a candidate but can still be fairly certain that they will win. That said, some more contemporary polls, including Emerson College Polling, have begun asking voters who they expect to win presidential elections \citep{emersoncollegepollingOctoberNationalPoll2020}. This raises an important question: is this type of information more valuable than traditional measures in predicting political outcomes?  If so, traditional polling approaches may be overlooking critical and informative data.

Capturing the degree of certainty of who people think will win (regardless of how they feel about the candidate), versus who they might want to win, is a key distinction worthy of closer examination. It’s also critical to examine the motivation by which participants would share their viewpoints, and whether asking questions through traditional polling methods provides enough motivation for individuals to share their views. While there are many ways to think about trying to obtain this information, one way to probe it could be by leveraging an old approach: betting. While the motivation behind betting has very complex origins (including reward systems in the brain), the belief that one is betting on what they think an outcome will be is the central premise and has held people’s interest through history. Indeed, ancient Egyptians bet on outcomes of games and sports such as fencing as long ago as 4000-3000 B.C. \citep{schwartzRollBonesHistory2006}. 

In the modern world, the advent of online betting markets, including Polymarket, provides an outlet for individuals to place bets on any number of future events, including presidential elections. These sources draw on a “Wisdom of Crowds” approach \citep{surowieckiWisdomCrowdsWhy2004}, which proposes that the average judgment of a large group of people can be more accurate than that of a smaller group, even if that small group is composed of individuals with more expertise or more vested interests in the matter \citep{davis-stoberWhenCrowdWise2014, larrickSocialPsychologyWisdom2012, sollStrategiesRevisingJudgment2009}. Wisdom of Crowds is thought to be widely observable across situations because its basic premise is rooted in maximizing the amount of information obtained, which reduces the impact of abhorrent sources and increases the representation of the sample, aiding validity \citep{budescuConfidenceAggregationOpinions2005}. The Wisdom of Crowds approach has been applied successfully for making public policy predictions \citep{morganUseAbuseExpert2014} as well as forecasting geopolitical decisions \citep{mellersPsychologicalStrategiesWinning2014}. In theory, this approach represents a much simpler method to evaluating outcomes compared to polling data, which often requires advanced statistical methods, like using weighting techniques \citep{zainoUSPresidentialElection2025, silverFINALSilverBulletin2024} that can introduce bias \citep{clintonPollingParadoxWhat2024, clintonLotStatePoll2024}.

\subsubsection{Polymarket}

Some previous research has investigated the accuracy of betting markets in predicting presidential elections both within and outside the U.S. \citep{caldeiraExpertJudgmentStatistical2004, eriksonArePoliticalMarkets2008, isotaloPredicting2016US2016, wolfersPredictionMarkets2004}; however, none have done so using Polymarket, the world’s current largest prediction market run on cryptocurrency. While other betting markets exist, including Iowa Electronic Markets (IEM), Kalshi, and PredictIt, they are limited in their predictive power for a number of reasons. First, the markets are much smaller than Polymarket, which had close to \$3.7 billion wagered on the 2024 presidential election. The markets of PredictIt and IEM are much smaller in size and did not collect state-level data. Thus, while these markets provide information about the general sentiment of the nation, they are not useful for tallying votes for the Electoral College, nor can they be used by candidates to inform their campaign strategies. Although the Kalshi website seems to indicate betting markets within swing states were collected, data collection didn’t begin until after October 2024, due to regulatory issues \citep{matthewsUSAppealsCourt2024}. 

A surface examination of the graphics provided by Polymarket suggests that betting provided remarkably accurate predictions of the U. S. presidential election outcome. According to posts on the social media app X, Polymarket had the odds of President Trump winning the 2024 election at 95\% before midnight of election day, several hours before the election was called in Trump’s favor via the Associated Press \citep{meirCalling2024Presidential2024, sundarPolymarketNailed20242024}. Despite initial interest in Polymarket’s accuracy in predicting the 2024 election, no comprehensive examination, especially alongside traditional pre-election polls, has been conducted to examine whether these betting markets, that have people bet on who they think will win (versus traditional polling methods, which typically report who people will be voting for), holds greater predictive power than polling. Here we provide the first comprehensive, systematic analysis to examine whether Polymarket data predicted the 2024 presidential election in a superior manner to traditional polling methods that have been in place for almost 100 years. Of note, we were cognizant that it would be critical to understand techniques to harden these markets ahead of future major events, such as the 2026 election, given the sensitive nature of the data and its vulnerability to market manipulation - even if more in depth examination of Polymarket supported prediction superiority for the 2024 presidential election.

\subsection{Current Study}

The current study examined whether betting market data provided insights that differed from traditional presidential polling. While a number of betting markets currently exist, in which individuals can place bets using cryptocurrencies, Polymarket is the largest \citep{sundarPolymarketNailed20242024}. Therefore, we utilized it as the market comparison versus polls in our analyses. At the outset, we were agnostic in our hypotheses, given that, on the face of it, a small segment of the population that uses cryptocurrency and furthermore places bets using it is likely not a broad representation of the United States population at large. Nonetheless, we put forth that if Polymarket predictions were indeed superior to polling, deep exploration, especially within swing states, would be needed to understand why such a phenomenon would be present and how this approach could be used to predict future events in other unexplored avenues. 

To examine the Polymarket versus polling data in the 2024 election, we took an overall approach of comparing the two via descriptive analyses and developing statistical models to fit and predict outcomes based on each data source. With the former, we were able to visually compare the accuracy and validity of the market data versus polling data. With the latter, we were able to identify key drivers underlying the dynamics of each data source and examine how trends in both, taken over various snapshots in time, predict outcomes leading up to Election Day. For all analyses, we utilized both national and state level data, with a focus on the seven swing states for the 2024 election (Arizona, Georgia, North Carolina, Pennsylvania, Michigan, Nevada, and Wisconsin).

\section{Methods}
\subsection{Data }

This study analyzed daily data on the probability of Donald Trump winning the 2024 U.S. presidential election using two primary sources: betting market data from Polymarket and public opinion polling data from multiple established polling aggregators. Both data sources were collected at the national and state levels. Polymarket data was downloaded via Kaggle \citep{andradePolymarket} and consisted of daily closing prices on markets tied to election outcomes, which were interpreted as implied probabilities of Trump winning the presidency (i.e., "full ballot"). Polling data was downloaded via the FiveThirtyEight website, which served to aggregate national polling data before it was shut down in March of 2025 \citep{dunbarPoliticalPollNews2025}.All available surveys on the corresponding day were included. When multiple polls were released on the same date for the same jurisdiction, the mean and standard deviation of the estimates were used to summarize the polling values for that day. A full list of the polling sources included can be referenced in Appendix A. All data were aligned by date, enabling a direct comparison between betting market predictions and polling-based estimates over time.

\subsection{Analyses}

We use a three-part analytical framework to evaluate the predictive power and underlying structures of the Polymarket betting markets and public opinion polls in the context of the 2024 U.S. presidential election. The first component involves a descriptive analysis of daily national and state-level data from Polymarket and polling sources, aimed at identifying trends, divergences, and responsiveness to major political events. The second component uses Bayesian Structural Time Series (BSTS) models to fit the Polymarket data and uncover key and potentially unknown drivers within the betting market. Lastly, the third part uses BSTS models to generate probabilistic forecasts from both Polymarket and polling data to assess their accuracy, stability, uncertainty, and evolving behavior as the election approaches.

\subsection{Descriptive Analysis}

The first phase of analysis involved a visual comparison of the Polymarket and polling data. Time series plots were generated from both data sources, highlighting trends at the national level and at the state-level for key swing states in the 2024 election (Arizona, Georgia, Nevada, North Carolina, Michigan, Pennsylvania, and Wisconsin). This visual comparison aimed to assess two main aspects of the data:

\begin{enumerate}
    \item 	\textbf{Accuracy}, focusing on how well each data source reflected developments and the eventual winner in the race.
    \item 	\textbf{Usability}, including aspects such as completeness of data, reactivity to major political events, and coherence in temporal trends
\end{enumerate}

\subsection{Model Fitting and Variable Importance}

For the second component of the analysis, BSTS models were trained on the general election Polymarket data to examine key drivers of the betting market. We used BSTS models because of their modular nature, allowing us to simultaneously decompose the time series into components representing long-term trends, short-term dynamics (e.g., autoregressive behavior), and the impact of external predictors \citep{brodersenInferringCausalImpact2015}. Additionally, the Bayesian nature of these models enabled us to naturally quantify bounds of uncertainty on parameter estimates \citep{brodersenInferringCausalImpact2015, scottPredictingPresentBayesian2013}. 

BSTS models are governed by general structural time series modeling, expressed by the following set of equations where yt denotes an observed value at time t \citep{scottPredictingPresentBayesian2013}: 

\begin{align}
y_t = Z_t^T\alpha_t + \epsilon_t ; \epsilon_t \sim N(0, H_t)\
\label{eq:1}
\end{align}

\begin{align}
\alpha_{t+1} = T_t\alpha_t + R_t\eta_t ; \eta_t \sim N(0, Q_t)\
\label{eq:2}
\end{align}

Here, Equation (1) is the observed equation because it links the observed data (yt) to an unobserved, latent state (t). Equation (2), the transition equation, describes how the latent state evolves over time \citep{scottPredictingPresentBayesian2013}. A wide variety of models, including ARIMA based models, are encompassed by this framework \citep{cowpertwait2009introductory}. 

A simple but effective form of this class of models is the local-level model with regressors, expressed by the following set of equations \citep{brodersenInferringCausalImpact2015}:

\begin{align}
y_t = \mu_t + \beta^Tx_t + \epsilon_t ; \epsilon_t \sim N(0, \sigma^2)\
\label{eq:3}
\end{align}

\begin{align}
\mu_{t+1} = \mu_t + \eta_t ; \eta_t \sim N(0, \tau^2)\
\label{eq:4}
\end{align}

We use this type of model as the baseline for fitting statistical models to the Polymarket general election data. Here, yt denotes the observed national Polymarket data at time t, and t captures the underlying temporal components (e.g., auto-regression) featured in the data. If 20, then t is constant, and the Polymarket dynamics are similar to Gaussian white noise fluctuating about this mean value. On the other hand, if 20, then the latent state evolves similar to a random walk through time where the next state depends solely on the most recent state. If both variance terms are positive, then there is a balance between impacts from past and recent data, determined by the ratio of the two terms. As such, by estimating these variance terms, we can gauge fundamental temporal mechanics governing the structure of the Polymarket betting data. 

Additionally, we also specify a vector of contemporaneous regressors (xt), which include Polymarket betting odds for the probability of Trump winning each of the 50 states (i.e., each state is a predictor). In doing so, we can see which states (e.g., key swing states) are most influential in shaping the overall Polymarket general election bets. Because the BSTS models feature spike-and-slab priors placed on regression coefficients (T), the models are encouraged to include only the most informative state-level predictors \citep{brodersenInferringCausalImpact2015}. 

It should be noted that we also explored different types of BSTS models, namely the \textit{semilocal linear trend} and the \textit{local linear trend} model, when fitting the data. These models incorporate increasingly more aggressive effects from temporal trends by including additional parameters in the latent state equation \citep{scottPredictingPresentBayesian2013}. 

\subsection{Forecasting and Dynamic Comparison}

Lastly, to simulate real-time predictive performances from each data source, BSTS models were also trained on national and state level data up to various intermediate time points prior to the election and then projected forward to election day. Here, only data from the time series being analyzed was included as inputs to the respective models. In other words, no regressors were used as model inputs because predictions for these regressors would have to be generated to make an overall model projection \citep{brodersenInferringCausalImpact2015, scottPredictingPresentBayesian2013}. Again, we use the local-level implementation of these models as the baseline for this analyses, and it can be expressed by the following set of equations:

\begin{align}
y_t = \mu_t + \epsilon_t ; \epsilon_t \sim N(0, \sigma^2)\
\label{eq:5}
\end{align}

\begin{align}
\mu_{t+1} = \mu_t + \eta_t ; \eta_t \sim N(0, \tau^2)\
\label{eq:6}
\end{align}

As seen, Equations (5) and (6) are identical to (3) and (4) respectively, except for the regression components featured in Equation (3). Setting the models up in this manner (i.e., no regressors) allowed us to more easily generate rolling forecasts from each data source. 

While the main goal of the previous analysis was to investigate dynamics underlying the data, the focus of this analysis is that of prediction. Namely, we examined the following key aspects of predicting values from each data source:
\begin{itemize}
    \item \textbf{Temporal stability}: assessing how predictions from each data source behaved as Election Day approached.
    \item \textbf{Event reactivity}: evaluating how major political events (e.g., an attempted assassination attempt, a change in party nominee) influenced predicted probabilities.
    \item \textbf{Uncertainty quantification}: comparing the width and informativeness of predictive intervals generated by the Bayesian models, which rely on the variance terms and temporal dynamics estimated by the models
\end{itemize}

All predictions were updated sequentially as new data became available, mirroring the accumulation of information during the lead-up to the election and providing an intuitive framework for comparing forecasts from poll-based and market-based data.

\section{Results}

Results from the descriptive analysis show that the Polymarket data was superior in correctly predicting Trump as the winner of the general election compared to national polling data (Figure 1). Additionally, the Polymarket data reacted more dynamically to events leading up to the election. For example, after the first assassination attempt on Trump in Pennsylvania on July 13, 2024, the Polymarket data exhibited increases in the probability of him winning. Similarly, after Kamala Harris entered the race on July 21st, Trump’s chances started to decrease. However, the polling data did not react similarly to these events and consistently hovered around a 45\% chance of Trump winning throughout the election timeline. 

\begin{figure}[htbp]
  \centering
  \includegraphics[width=1\textwidth]{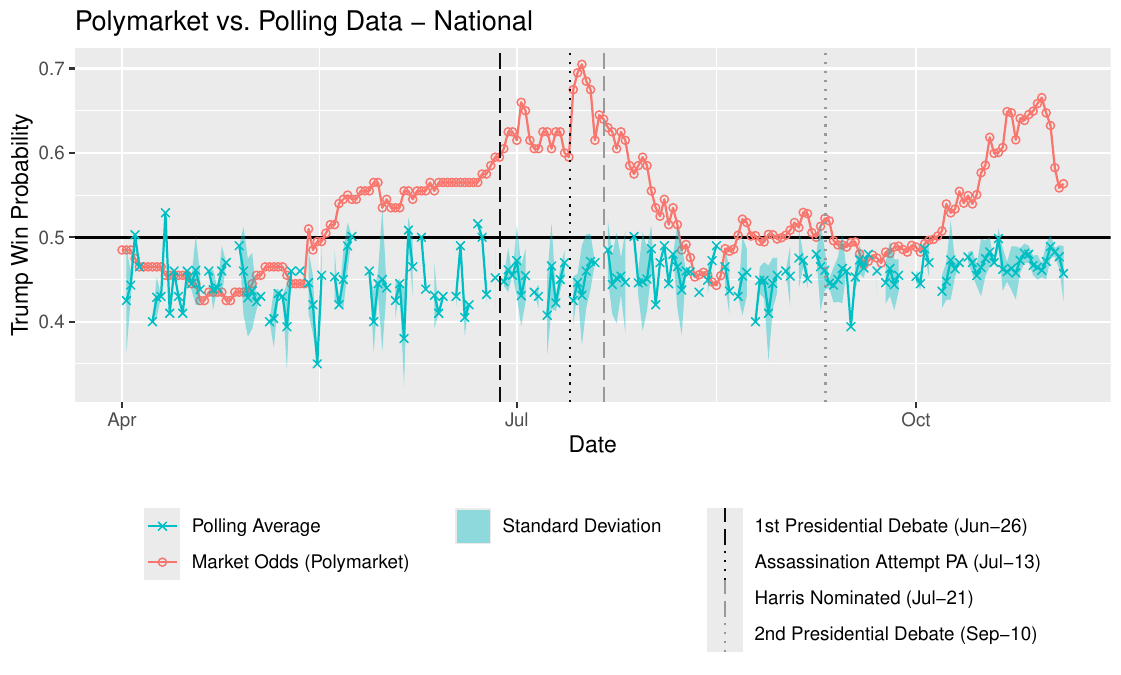}
  \caption{Polymarket Versus Polling Data (National) – Trump was the winner of the general election (2024)}
\end{figure}

The state-level comparisons (Figure 2) reflect much of the same findings as those from the general election. The Polymarket data was also superior at predicting outcomes for most of the key swing states, including Arizona, Georgia, North Carolina, and Nevada. Although Pennsylvania is slightly less clear, the chart still mostly favors the Polymarket data, especially as the election drew closer. For Wisconsin and Michigan, the Polymarket data appear to oscillate about mean values that also interestingly align with the polling averages. For these two states, perhaps the markets were truly suggesting that the corresponding outcomes were too close to call ahead of time. 

\begin{figure}[htbp]
  \centering
  \includegraphics[width=1\textwidth]{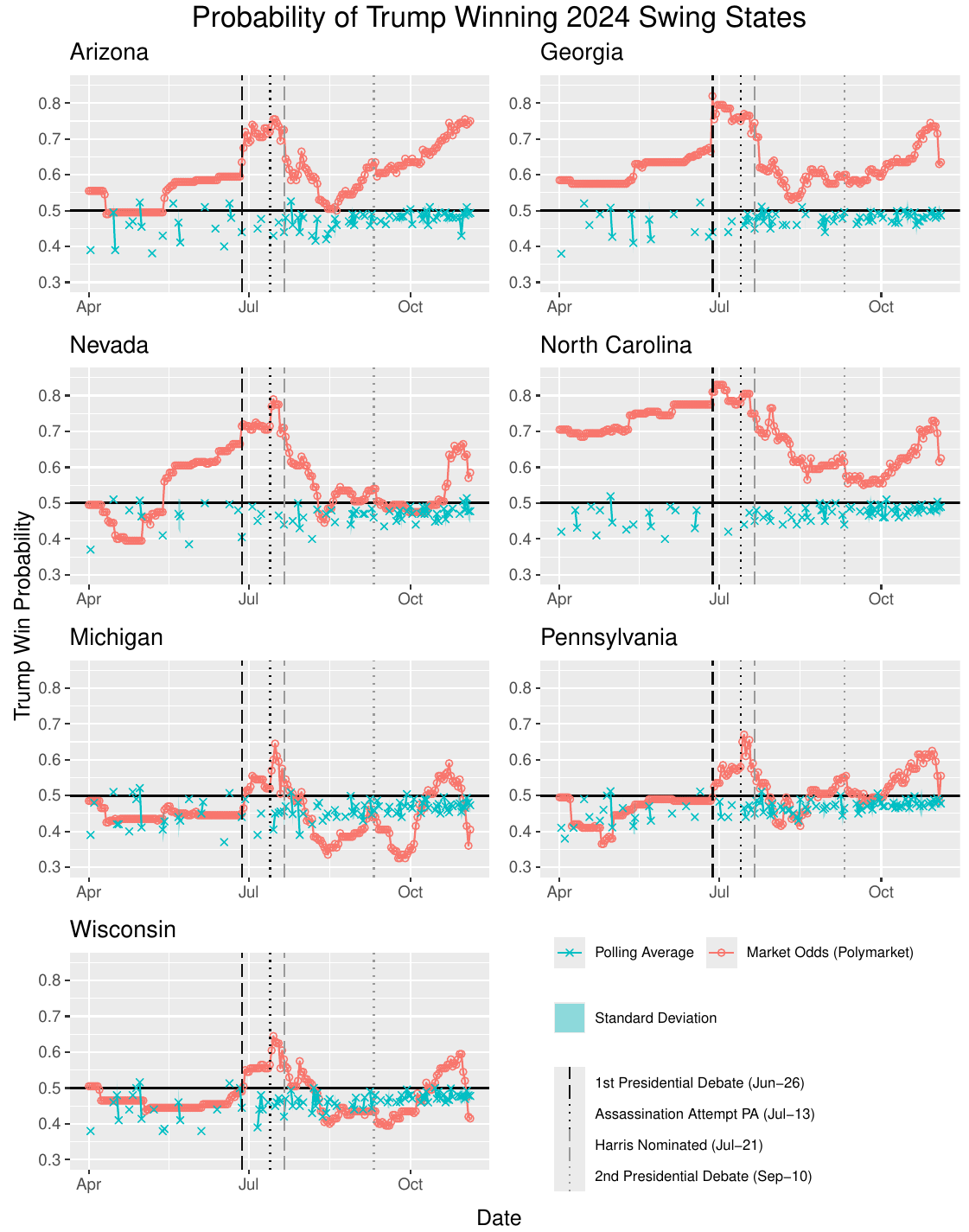}
  \caption{Polymarket Versus Polling Data (State) – Trump was the winner for each of these states (2024)}
\end{figure}

From a practical standpoint, the Polymarket data were also more robust and complete, due to the fact the data was recorded daily. In contrast, the polling data was collected via intermittent sampling, thus resulting in more missingness. This fact is even more evident in the state-level data (Figure 2). When modeling and analyzing the data in more detail, having a mostly complete set of data is highly beneficial. 

The BSTS models were able to accurately fit the general election Polymarket data (e.g., Figure 3a). Additionally, the 95\% credible intervals are tight, indicating posterior estimates for the parameters are precise. As seen in Figure 3b, two of the most notable swing states, Pennsylvania and Michigan, are consistently featured as important predictors. As discussed, spike-and-slab priors are placed on regressor coefficients to encourage a sparse feature set, so only the most influential regressors are consistently featured in the models \citep{brodersenInferringCausalImpact2015, scottPredictingPresentBayesian2013}. In other words, these two swing states played a significant role in shaping the overall Polymarket bets for the general election. 

\begin{figure}[htbp]
  \centering
  \includegraphics[width=1\textwidth]{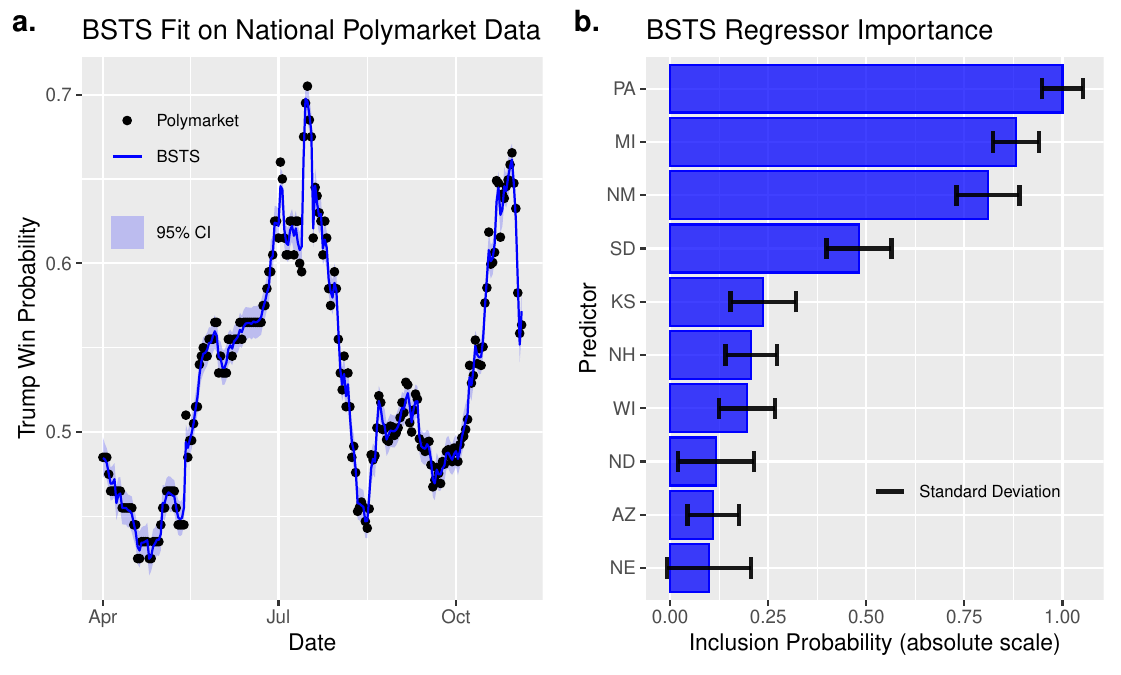}
  \caption{BSTS Local-Level Model on National Polymarket Data}
\end{figure}

It should be noted that the results in Figure 3 pertain to the local-level BSTS model, specified in Equations (3) and (4). The same analysis was repeated for the semilocal linear trend and the local linear trend model, but the local-level model performed marginally better than these two. Additionally, insights stemming from the other models were largely the same as that of the local-level model. As such, results from the semilocal linear trend and local linear trend models are not included in the main manuscript. However, they are included in the Supplementary Materials, along with the code and data necessary to reproduce all analyses featured in the paper. 

Lastly, results from the predictive models are shown in Figure 4. Here, BSTS models were trained on the Polymarket and polling data up to specified time points for the general election based on Equations (5) and (6). Then, predictions were extrapolated to Election Day. Similar sets of predictions were also made for each swing state (Supplementary Materials). Findings largely aligned with the findings from the national predictive models, with some subtle distinctions (noted in the Discussion section).

As seen in Figure 4, forecasts from the Polymarket data appear to be less precise than the polling data. The 95\% predictive interval bands from the Polymarket model predictions fan out through time while those from the polling data remain consistent. This finding is due to the underlying dynamics of the two data sources that the BSTS models captured. The Polymarket data exhibit more of a random-walk like time series characteristic of financial markets (i.e., has a higher degree of auto-regression). The resulting estimates for the variance parameters included in the model corroborate this notion (2). Conversely, the polling data appear to be more characteristic of Gaussian noise distributed over a mean value, again corroborated by resulting parameter estimates (0 and >0). 

\begin{figure}[htbp]
  \centering
  \includegraphics[width=1\textwidth]{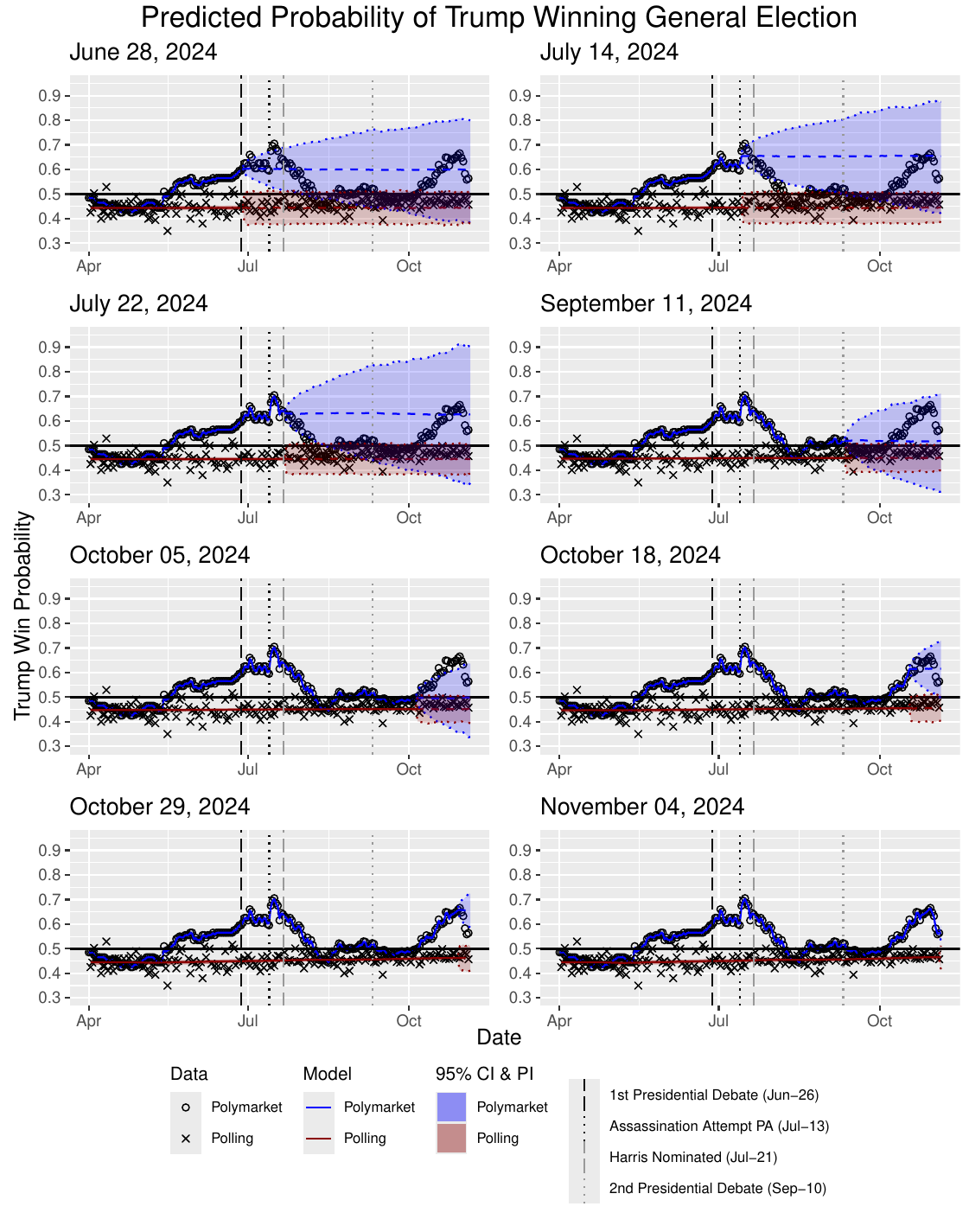}
  \caption{Model Predictions for Polymarket and Polling Data }
\end{figure}

However, even with the increased uncertainty due to the random-walk-like structure of the Polymarket data, overall predictions are more accurate than those of the polling data. The mean posterior prediction across time mostly favors Trump, versus that of the polling data always favoring Harris. Moreover, most posterior predictions from the Polymarket data contain the correct prediction (i.e., most of the blue funnel in Figure 4 falls above the 50\% decision boundary in favor of Trump). Conversely, almost none of the predictions from the polling data favor Trump (i.e., almost all polling uncertainty prediction intervals fall below the 50\% probability line).  As election day draws nearer, the forecasts from the Polymarket models become more precise (i.e., they extrapolate the random-walk over a shorter duration). By mid-October, almost all predictions from the Polymarket data have Trump winning while at the same time, almost all predictions from the polling data favor Harris. 

Lastly, the Polymarket forecasts respond much more dynamically to events leading up to the election than those from the polling data. After the first assassination attempt in Pennsylvania, Trump’s odds of winning are intuitively predicted to be higher in the Polymarket forecasts. However, those of the polling data remain unchanged. Analogous trends occur when Harris becomes the nominee and performs well during the second Presidential debate. Here, we see the Polymarket predictions reflect a sentiment shift toward Harris, but predictions for the polling data remain largely unchanged. 

\section{Discussion}

The current study compared the probability of Donald Trump winning the 2024 U.S. presidential election from two primary sources: pre-election polling data and betting market data from one of the largest betting markets, Polymarket. We examined the data both descriptively (i.e., to explore data trends now that the event has passed) and using predictive analytics. The predictive analysis allowed us to examine in “real time” how well the prior findings predicted the election outcomes as the timeline progressed (i.e., from April 2024 through election day on November 4th, 2024). Overall, findings suggest that Polymarket was superior to polls in predicting the 2024 presidential election outcomes. Below we elaborate on findings from our descriptive and predictive analyses and discuss their implications for future polling data and elections.

\subsection{Descriptive Results}

Descriptive analyses of the national Polymarket data indicate that, at all but two time points (in May and September of 2024), President Trump was favored to win the 2024 election. This stands in stark contrast to the polling data, which show a much tighter race between the two candidates over the full April to November timeline. Further, this effect within the Polymarket data becomes increasingly evident in October of 2024, after which Trump’s probability of winning peaks at nearly 67\% and continues to stay above 55\% until election day. Also starting in the month of October, the Polymarket data is no longer overlapping with standard deviation estimates of the polling data. Nonetheless, it is important to note that the emergence of the trend towards Trump coincides with large bets being placed by one person across many accounts \citep{osipovichMystery30Million2024a}, which have prompted speculations of market manipulation \citep{saulTrumpsPolymarketOdds2024, silverWhatsTrumpsSurge2024, sundarPolymarketNailed20242024}. The degree of this manipulation and its influence on the market/voting outcomes is debated \citep{corbaPolymarketPlansUS2024, morrowHowPredictionMarkets2024} and will need further examination in future studies.

When examining the swing states, findings demonstrate that Polymarket consistently shows Trump as the winner for Arizona, Georgia, and North Carolina – both before and after Harris became the nominee. For Nevada and Pennsylvania, Polymarket data show Trump as the clear winner from mid-October onward, except for the early morning hours of November 3rd, 2024.  Finally, data for Wisconsin and Michigan reflect that polling data may be potentially superior to Polymarket data. While all states show a negative drop in Trump’s winning probability on November 3rd, 2024, the states of Michigan and Wisconsin show the largest dips in winning likelihood. Thus, it may be that for the states of Michigan and Wisconsin, the polls were accurate in suggesting their outcomes were too close to call. Future investigations into why the Polymarket data from the states of Wisconsin and Michigan show unique patterns compared to the other swing states, is necessary.
Overall, across the swing states, the Polymarket data clearly show Trump is more likely to win the presidential election for five out of the seven of them. Further, descriptive analyses of the Polymarket data suggest the swing states of Georgia and North Carolina were never actually battleground states to begin with. Indeed, the Polymarket data indicate the likelihood of Trump winning the 2024 election in Georgia and North Carolina rarely drops below 55\%, often staying between 60-80\%, while the polling data shows estimates that hover around 40-50\%. A similar pattern of results is also present between Arizona and Nevada; however, the trend in the Polymarket data becomes apparent after June of 2024, after which the polling data continue to suggest it is too close to call.

Overall, the Polymarket gives unique insight into voting preferences within the swing states beyond that of the polls. Indeed, the Polymarket data suggest that leading up to the election, Trump could have spent less money and time campaigning in the typical battleground states of Arizona, North Carolina, and Georgia and instead reallocated those resources to the states of Michigan and Wisconsin that were still neck and neck. For Harris’s team, the Polymarket data indicate she needed to continue campaigning and likely increase her campaigning efforts in the battleground states, as the data showed she was behind Donald Trump more than the polling data suggested. While the above findings are intriguing, it is worth noting that the above interpretation involves data that are only descriptive in nature. Therefore, they do not allow us to compare how well the national Polymarket or polling data predicted the outcome of the 2024 election.

\subsection{Predictive Results}
To address this limitation, we ran predictive analyses using BSTS approaches, which allowed us to predict outcomes from prior data. We examined the predictive power of the Polymarket and polling findings individually at 8 time points (from June to November) using 95 percent predictive intervals. Starting around October 18th, 2024, the Polymarket data not only predicts Trump as the winner via its mean line, but the 95 percent predictive intervals no longer cross below the 50/50 chance line of Trump winning – therefore, even taking uncertainty into account, Polymarket data predicts Trump will be the winner of the 2024 election as early as mid-October of 2024. In contrast, the mean of the polling data never showed Trump as the winner; furthermore, the predictive interval of the polling data always hovered below and above 0.5, therefore never firmly predicting either candidate as a clear winner. The only exception to this trend was on election day, for which the polling data incorrectly predicts Harris as the winner.

The predictive analyses also allowed us to compare whether the two different approaches to predicting election outcomes, Polymarket versus polling data, showed clear divergence (e.g., by examining where the 95\% predictive intervals no longer overlapped). Results of these analyses suggest that starting around October 18th, 2024, the two approaches, with their unique data sources, were predicting the 2024 election outcome differently. While this result suggests that the two approaches are distinct, what is behind this phenomenon is unclear. The most obvious explanation may be that the individuals placing bets in Polymarket represent a different population/demographic than what is represented by the polling data; however, the demographic characteristics of Polymarket users is relatively unknown.  It is notable that the Polymarket platform requires users to use cryptocurrency, which estimates suggest only 17-20\% of Americans own \citep{matosSurveyReveals12025, murphyJDVanceSays2025}. Further, because of a CFTC settlement, Polymarket bets were also supposed to be restricted to individuals outside of the United States \citep{cftcCFTCOrdersEventBased2022}, whereas polling data for United States presidential elections, including the data used in this analysis, are typically limited to individuals within the United States.

When examining the swing states, most of the Polymarket findings were clear in predicting Trump as the winner by mid-to-late October. This is true for the states of Arizona, North Carolina, Georgia, Pennsylvania, and Nevada, for which the 95 percent predictive intervals for Trump winning the election stay above 50\%. In contrast, the predictive intervals of the polling data for these same states always hovered below and above chance levels. When examining the means only, Polymarket data for Arizona, North Carolina, Georgia, Nevada, and Pennsylvania predicted Trump as the winner by September or earlier. When examining the polling data, means largely suggested Harris as the potential winner of these states; however, the predictive intervals continued to hover below and above chance at all timepoints.

In terms of comparing whether/when the two approaches showed clear divergence within the swing states in terms of non-overlapping predictive intervals, results showed a similar pattern as the national data (see Supplementary material), and demonstrated that the Polymarket data for Arizona, North Carolina, and Georgia were distinct from polling data from October 18th, 2024, onward. Results for Nevada and Pennsylvania indicate statistical differences emerged by October 29th (almost on October 22nd, 2024, for Pennsylvania). Again, these results suggest the two approaches to data collection were capturing different underlying issues and/or themes for these five swing states. As seen in the descriptive data, predictions for the states of Michigan and Wisconsin are always unclear, regardless of which data set is examined. Although the Polymarket data for both states had means favoring Trump as the winner, the predictive intervals for both the polling and Polymarket data in these two states indicate the models were performing largely at chance levels. In keeping with this theme, the states of Michigan and Wisconsin did not show ever show divergence/non-overlapping predictive intervals in Polymarket versus polling data. Again, investigations into why these states have unique descriptive and predictive patterns compared to the other swing states is worthy of further investigation.

\subsection{Limitations}
In general, results underscore the accuracy of the Polymarket data compared to the polling data in predicting the outcome of the 2024 presidential election; however, there are several limitations that need to be considered. First, the betting market data are from a single source, while the polling data are aggregated across many polls. Therefore, we were unable to generate means and standard deviations of the Polymarket data that are comparable to that of the national polling data. Whether the patterns observed in the Polymarket data extend to other betting markets, including IEM, Kalshi, and PredictIt, remains to be seen. An initial inspection suggests some of the trends are similar across many of the markets (but not all, e.g., IEM predicted Harris as the winner), despite the data from the other betting markets being less robust and more incomplete. A more detailed investigation is necessary to determine whether the effects presented in this study can be generalized across all betting markets. Perhaps as betting markets become more prevalent, especially in the United States, this will allow for more robust examination of the value of betting markets for predicting future event outcomes.

Second, it is not only unclear how many individuals were betting on Polymarket, but also the demographic characteristics of those individuals is unknown. This issue is especially relevant regarding the singular individual who was confirmed to be placing large bets across multiple accounts \citep{osipovichMystery30Million2024a}. Beyond this specific limitation, Polymarket was also not supposed to be open to bettors within the United States \citep{cftcCFTCOrdersEventBased2022}. Thus, the sample should have been very unique from that of the polling data, which was done with American citizens; however, sources indicate Americans were capitalizing on the anonymity of the website and betting via VPNs \citep{beyoudUSTradersFlock2024, sundarPolymarketNailed20242024}. If true, and a large sample of Americans were participating in Polymarket, the representativeness of the sample is likely still limited due to the fact that the platform is based in cryptocurrency, since only about 17-20\% of Americans have cryptocurrency \citep{matosSurveyReveals12025, murphyJDVanceSays2025}. Yet, this demographic information, if obtained, may prove informative, specifically for groups that the polls may not often accurately capture, like Republicans, who tend to have lower response rates to polls \citep{kennedyKeyThingsKnow2024, cramerPoliticsResentmentRural2016}. Republicans, specifically Republican men, are more likely to report higher rates of crypto ownership \citep{blackstone2025CryptocurrencyAdoption2025, deviseRepublicansWereCrypto2025}. Therefore, it is possible that betting markets may simply be better predictors of election outcomes because they are capturing more of a demographic that polls often miss.

\subsection{Conclusions and Future Directions}

There are several notable differences between the Polymarket and polling data that are worth mentioning. First, Polymarket data relies on a Wisdom of Crowds approach to predicting event outcomes. This framework suggests the average judgment of a large group of people is accurate in predicting binary event outcomes, and although it is similar to that of the polls, it is much simpler in its approach; Polymarket projections do not involve weighting or other statistical factors that can bias results in polling \citep{clintonPollingParadoxWhat2024, clintonLotStatePoll2024}. Second, the data from Polymarket involves some risk for the individual, as people are actively putting money down on an outcome, whereas polling data largely captures people’s opinions and/or intent. How this difference affects people’s choices and general group consensus accuracy merits future investigation. Finally, the betting markets produce much more granular time series data (by day, by hour, and even some by the minute) whereas polls are conducted over a shorter time frame, usually over a few days \citep{gramlichHowPeopleUS2024}. Thus, the online betting market data captures unique insight into people’s thoughts on events as they happen in real time and allows for the examination of how these insights change in response to real-world events, such as Trump’s assassination attempt, Biden’s debate performance, or Harris’s replacement of Biden as the Democratic nominee. 

It is important to highlight the strong possibility that by exposing the predictive power of the Polymarket data for the 2024 election, we may be breaking the phenomenon of interest going forward. It’s unclear whether the release of this information will be used to influence voting behavior in the future. Indeed, some have argued against the use of data visualization during pre-election polling as it can impact voting behavior \citep{zainoUSPresidentialElection2025, bergResultsDozenYears2008}, and although this situation is not unique to betting markets alone, the fact that people have wagered money on these outcomes likely only increases the chances of this happening. To limit this potential problem, some have suggested blackout dates for both polls and betting markets leading up to election day \citep{rossBanPollsUK2024}, but it is unclear whether doing so would adequately address the full scope of the problem and address the real issue of how to harden markets.

A second problem regarding online Polymarket data is that there is limited guidance for how the public should use and understand it and its projections. Without this guidance, it’s possible the markets may come to be relied on as fact, when they are only wagered projections. This problem is further compounded by the fact that Polymarket and other online betting markets are currently vulnerable to market manipulation, tampering, and gamification. Speculation of online prediction market manipulation is not new \citep{cassidyWhatKilledIntrade2013, saamaPoliticalBiasPrediction2025}. During the 2024 presidential election, Polymarket had accusations of wash trading, a type of market manipulation in which shares are repeatedly bought and sold to boost volume and activity \citep{schwartzExclusiveElectionBetting2024}. There were also accusations of market tampering due to the large payments on Polymarket that were traced to one person across multiple accounts \citep{corbaPolymarketPlansUS2024, morrowHowPredictionMarkets2024, saulTrumpsPolymarketOdds2024, silverWhatsTrumpsSurge2024, sundarPolymarketNailed20242024}. These behaviors result in misleading market increases that can be falsely interpreted as market growth and, in the case of presidential markets, can lead to increases in market wagering, campaign donations, and/or voting behavior by creating a false sense of popularity \citep{imisikerWashTradesStock2018}. As such, additional research into how to best harden these markets (i.e., through anti-manipulation, identify verification, transparency, and auditability tools), so that they are not vulnerable to these threats, is necessary. 

In conclusion, the current study was the first to systematically compare traditional polling data to betting market data from one of the largest betting markets, Polymarket. This type of analysis was motivated in part from the inaccuracies of polling data in the past several presidential elections \citep{kennedyHowPublicPolling2023, kennedyKeyThingsKnow2024}, as well as the increase in online betting markets for predicting binary events (e.g., Polymarket and Kalshi). Descriptive and predictive results indicate national Polymarket data was superior to that of the polling data in predicting Trump to be the winner of the 2024 presidential election. Within the swing states specifically, examination of the Polymarket data provided unique insight into voting behavior beyond that which was available from the polls. 

These results suggest new Wisdom of Crowds methods could be employed to predict presidential elections; however, whether these findings replicate across other elections, and, furthermore, other world events, remain to be seen. We understand that exposing this phenomenon will have far-reaching implications – we are creating a new line of industry and/or research focused on determining whether this type of telemetry can be useful in predicting other events. This initial research project was done in an effort to prompt further research into how this real-word data approach can be transferred to new disciplines, as well as to expose the current vulnerability of online betting markets, like Polymarket. Interdisciplinary investigations involving political and computer scientists, and beyond, are necessary to determine how best to create and fortify online betting markets so that they can be leveraged responsibly. Continued research and learning is necessary to provide insight into the aforementioned limitations and to determine what these findings mean going forward into future elections and beyond.

\bibliographystyle{plainnat}
\bibliography{References}

\clearpage

\appendix
\section*{Appendix A: Pollster Names}

\begin{multicols}{2}
\begin{itemize}
\item 1892 Polling
\item 1983 Labs
\item 3W Insights
\item Abacus Data
\item ABC/Washington Post
\item ActiVote
\item Alaska Survey Research
\item American Pulse
\item Angus Reid
\item Arc Insights
\item ARW Strategies
\item AtlasIntel
\item Axis Research
\item Beacon/Shaw
\item Bendixen \& Amandi International
\item Benenson
\item Benenson Strategy Group/GS Strategy Group
\item Big Data Poll
\item Big Village
\item BK Strategies
\item Blueprint Polling
\item Bowling Green State University/YouGov
\item Bullfinch
\item Capitol Weekly
\item Causeway Solutions
\item CES / YouGov
\item Change Research
\item Cherry Communications
\item Chism Strategies
\item Christopher Newport U.
\item Civiqs
\item Claflin University
\item Clarity
\item CNN/SSRS
\item co/efficient
\item Cor Services, Inc.
\item Cor Strategies
\item CWS Research
\item Cygnal
\item Cygnal/Aspect Strategic
\item Dan Jones
\item Dartmouth Poll
\item Data for Progress
\item Data Orbital
\item Data Viewpoint
\item Deltapoll
\item DHM Research
\item Differentiators
\item Digital Research
\item East Carolina University
\item Echelon Insights
\item Echelon Insights/GBAO
\item Elon U.
\item Elway
\item Embold Research
\item Emerson
\item EPIC/MRA
\item Fabrizio
\item Fabrizio Ward
\item Fabrizio/David Binder Research
\item Fabrizio/GBAO
\item Fabrizio/Impact
\item Fabrizio/McLaughlin
\item Fairleigh Dickinson
\item Faucheux Strategies
\item Fleming \& Associates
\item Florida Atlantic University
\item Florida Atlantic University/Mainstreet Research
\item Focaldata
\item Fort Hays State University
\item Franklin and Marshall College
\item Gateway Political Strategies
\item Glengariff Group Inc.
\item Global Strategy Group
\item Global Strategy Group/North Star Opinion Research
\item Gonzales Research \& Media Services
\item GQR
\item Gravis Marketing (post 2020)
\item GS Strategy Group
\item Harris Poll
\item HarrisX
\item HarrisX/Harris Poll
\item Hart Research Associates
\item Hart/POS
\item Hendrix College
\item High Point University
\item HighGround
\item Hoffman Research
\item Hunt Research
\item Impact Research
\item Innovative Research Group
\item InsiderAdvantage
\item Ipsos
\item Iron Light
\item J.L. Partners
\item John Zogby Strategies
\item KAConsulting LLC
\item Kaiser Family Foundation
\item Kaplan Strategies
\item Keating Research
\item Lake Research
\item Landmark Communications
\item Leger
\item Let's Preserve the American Dream
\item Lord Ashcroft Polls
\item Manhattan Institute
\item Marist
\item Marquette Law School
\item Mason-Dixon
\item MassINC Polling Group
\item McCourtney Institute/YouGov
\item McLaughlin
\item Meeting Street Insights
\item Meredith College
\item Metropolitan Research
\item Miami University (Ohio)
\item Michigan State University/YouGov
\item Mitchell
\item Monmouth
\item Morning Consult
\item MRG (Marketing Resource Group)
\item MSU - Billings
\item Muhlenberg
\item Murphy Nasica \& Associates
\item National Public Affairs
\item National Research
\item Navigator
\item New Bridge Strategy/Aspect Strategic
\item NewsNation/Decision Desk HQ
\item Noble Predictive Insights
\item NORC
\item North Star Opinion Research
\item OH Predictive Insights / MBQF
\item Ohio Northern University Institute for Civics and Public Policy
\item OnMessage Inc.
\item Opinion Diagnostics
\item Outward Intelligence
\item P2 Insights
\item Pan Atlantic Research
\item Paradigm
\item Patriot Polling
\item Peak Insights
\item PEM Management Corporation
\item Pew
\item PPIC
\item PPP
\item Praecones Analytica
\item Premise
\item Prime Group
\item Probolsky Research
\item Project Home Fire
\item PRRI
\item Public Opinion Strategies
\item PureSpectrum
\item Quantus Insights
\item Quantus Polls and News
\item Quinnipiac
\item RABA Research
\item Rasmussen
\item Redfield \& Wilton Strategies
\item Remington
\item Research \& Polling
\item Research America
\item Research Co.
\item RG Strategies
\item RMG Research
\item Roanoke College
\item Rutgers-Eagleton
\item Schoen Cooperman
\item Selzer
\item Siena
\item Siena/NYT
\item Slingshot Strategies
\item SoCal Research
\item SoCal Strategies
\item SoonerPoll.com
\item Split Ticket
\item Split Ticket/Data for Progress
\item Spry Strategies
\item SSRS
\item St. Anselm
\item St. Pete Polls
\item Stetson University Center for Public Opinion Research
\item Strategies 360
\item Suffolk
\item Survation
\item Survey Center on American Life
\item SurveyMonkey
\item SurveyUSA
\item SurveyUSA/High Point University
\item Susquehanna
\item Target Insyght
\item Targoz Market Research
\item Tarrance
\item Texas Hispanic Policy Foundation
\item Texas Lyceum
\item The Citadel
\item The Political Matrix/The Listener Group
\item The Tyson Group
\item The Washington Post
\item TIPP
\item Torchlight Strategies
\item Trafalgar Group
\item Trafalgar Group/InsiderAdvantage
\item Tufts
\item Tulchin Research
\item U. Arizona/TrueDot
\item U. Georgia SPIA
\item U. Houston
\item U. Massachusetts - Lowell
\item U. New Hampshire
\item U. North Florida
\item U. Rhode Island
\item UC Berkeley
\item UMass Amherst/YouGov
\item UMBC
\item University of Akron
\item University of Delaware
\item University of Houston/Texas Southern University
\item University of Maryland/Washington Post
\item University of Maryland/YouGov
\item University of Massachusetts Lowell/YouGov
\item University of Texas at Tyler
\item University of Wyoming
\item UpOne Insights/BSG
\item USC Dornsife/CSU Long Beach/Cal Poly Pomona
\item VCreek/AMG
\item Victory Insights
\item Virginia Commonwealth U.
\item VoteTXT
\item Washington Post/George Mason University
\item Wick
\item Winthrop U.
\item WPAi
\item Yale Youth Poll
\item YouGov
\item YouGov Blue
\item YouGov/Center for Working Class Politics
\item YouGov/SNF Agora
\item Z to A Research
\item Zogby

\end{itemize}
\end{multicols}

\end{document}